\documentstyle[12pt]{article}
\begin{document}
\tolerance=5000
\def\be{\begin{equation}}
\def\ee{\end{equation}}
\def\bea{\begin{eqnarray}}
\def\eea{\end{eqnarray}}
\def\nn{\nonumber \\}
\def\cF{{\cal F}}
\def\det{{\rm det\,}}
\def\Tr{{\rm Tr\,}}
\def\e{{\rm e}}

\  \hfill 
\begin{minipage}{3.5cm}
NDA-FP-32 \\
May 1997 \\
\end{minipage}

\ 

\vfill

\begin{center}

{\Large\bf Effective Potential for D-brane in Constant Electromagnetic Field}

\vfill

{\large\sc Shin'ichi NOJIRI}\footnote{
e-mail : nojiri@cc.nda.ac.jp}
and
{\large\sc Sergei D. ODINTSOV$^{\spadesuit}$}\footnote{
e-mail : odintsov@quantum.univalle.edu.co or sergei@ecm.ub.es}

\vfill

{\large\sl Department of Mathematics and Physics \\
National Defence Academy \\
Hashirimizu Yokosuka 239, JAPAN}

{\large\sl $\spadesuit$ Dep.de Fisica \\
Universidad del Valle \\
AA25360, Cali, COLUMBIA \\
and \\
Tomsk Pedagogical Univer. \\
634041 Tomsk, RUSSIA}

\vfill

{\bf ABSTRACT}

\end{center}

We discuss the one-loop effective potential 
and static (large $d$)
potential for toroidal D-brane described by DBI-action 
in constant
magnetic and in constant electric fields. Explicit 
calculation is
done for membrane case ($p=2$) for both types of 
external fields
and in case of static potential for an arbitrary $p$.
In the case of one-loop potential it is found that 
the presence of
magnetic background may stabilize D-brane giving the possibility
for non-pointlike ground state configuration. On the same time, constant
electrical field acts against stabilization and the correspondent
one-loop potential is unbounded from below.
 The properties of static potential
which also has stable minimum are described in detail. The back-reaction
of quantum gauge fields to one-loop potential is also evaluated.

\vfill


\newpage

\section{Introduction}

There was recently much of interest in the study 
of the dynamics
of $D=11$ $M$-theory \cite{1}. One of the key 
ingredients of $M$-theory
is D-brane \cite{2,3} which describes the 
non-perturbative dynamics
of string theory \cite{4} and gives new insight 
into understanding
of stringy black holes \cite{5}.

D-brane theories are described by Born-Infeld(BI)-type 
actions where
the world volume vector field naturally appears. 
Such theories
have been the subject of much recent work \cite{6}. 
Note that BI-
type effective action of open string theory has 
been discussed
in refs.\cite{7}.
  
It is well-known that some physical properties 
of the extended objects 
in the quantum regime may be gleaned from 
a study of the effective action
for various static $p$-brane configurations. 
In this way one studies
the effective action in one-loop or in large $d$ 
approximation. For
string case, large $d$ approximation has been developed 
in ref.\cite{8}
as systematic expansion for the effective action in 
powers of $1/d$.
The static potential may be then obtained by studying 
the saddle point
equations for the leading order term. Such program has 
been realized
also for rigid string \cite{9}, rigid string at 
non-zero temperature
\cite{10} (where it is expected to be helpful in 
drawing the relations
between QCD and string theory), bosonic membrane 
\cite{11,12} and bosonic
$p$-brane \cite{12}.

The purpose of present work is to investigate the 
one-loop and static
potentials for D-brane in the background gauge fields. 
We start such
calculation in the next section for situation when 
external electromagnetic
fields are purely classical. The one-loop potential 
in the constant magnetic
and in the constant electric field is found for 
$p=2$ (membrane). We show
that unlike the case without electromagnetic field there 
exists non-trivial
minimum of effective potential corresponding to the non-pointlike ground state.
The static potential in the same background is also 
calculated and its properties are discussed. 
Third section is devoted to the study of back reaction
of quantum gauge fields to effective potential. 
We calculate the one-loop
potential in the same background with additional 
contribution due to quantum gauge fields but it is found that the 
qualitative natures of the obtained one-loop potential are not
changed.
Some remarks and outline are given in Discussion.

\section{One-loop and static potential for D-brane in constant 
electromagnetic field.}

The D-brane is described by the 
Dirac-Born-Infeld-type action \cite{6}
\be
\label{DBIS}
S_D=k\int_0^T d\zeta_0 \int d^p\zeta \e^{-\phi(X)}
\left[\det \left((G_{\mu\nu}+B_{\mu\nu})
\partial_i X^\mu\partial_j X^\nu+F_{ij}\right)
\right]^{1 \over 2}\ .
\ee
Here $X^\mu$'s are the coordinates of D-brane 
($\mu, \nu = 0,1,\cdots,d-1$), 
$\zeta^i$'s are the coordinates on the D-brane world 
sheet ($i,j=0,1,\cdots, p$),
$G_{\mu\nu}$ is the metric of the space-time, 
$B_{\mu\nu}$ is the anti-symmetric tensor and 
$F_{ij}$ is the electromagnetic field strength on 
the D-brane world sheet:
\be
\label{fstrgth}
F_{ij}=\partial_i A_j - \partial_j A_i \ .
\ee
We are now interested in the stability of the D-brane 
and study the effective potential.
The effective potentials of $p$-brane were 
studied in \cite{12}.
In this paper, we investigate the case that $F_{ij}$ (or 
$B_{ij}$) has nontrivial vacuum expectation value. 

If we choose the following background gauge choice
\be
\label{ggch}
X^i=R_i \zeta^i \ \ (R_0=1)\ \ \ \ i=0,1,\cdots,p
\ee
and impose the periodic boundary conditions corresponding to
the toroidal D-brane
\bea
\label{pbc}
&& X^m(\zeta^0, \zeta^1, \zeta^2, \cdots, \zeta^p)
 =X^m(\zeta^0+T, \zeta^1, \zeta^2, \cdots, \zeta^p) \nn
&& =X^m(\zeta^0, \zeta^1+1, \zeta^2, \cdots, \zeta^p) 
 =X^m(\zeta^0, \zeta^1, \zeta^2+1, \cdots, \zeta^p) \nn
&& = \cdots =X^m(\zeta^0, \zeta^1, \zeta^2, \cdots, \zeta^p+1) \nn
&& (m=p+1,p+2,\cdots,d-1)\ ,
\eea
we obtain the following gauge-fixed action, 
\bea
\label{gfaction}
S_D&=&k\int_0^T d\zeta_0 \int d^p\zeta  \e^{-\phi(X)} \nn
&& \times
\left[\det \left(\hat G_{ij} + \hat G_{im}\partial_j X^m
+\hat G_{mj}\partial_i X^m
+G_{mn} \partial_i X^m \partial_j X^n \right)
\right]^{1 \over 2}\ .
\eea
Here
\bea
\label{G0}
\hat G_{ij}&=&R_iR_j(G_{ij}+B_{ij})+F_{ij} \nn
\hat G_{im}&=&R_i(G_{im}+B_{im}) \nn
\hat G_{im}&=&R_i(G_{im}+B_{im}) \nn
\hat G_{mn}&=&G_{mn}+B_{mn}\ .
\eea

Note that in the above gauge there are no Faddeev-Popov ghosts.
For simplicity, we only consider the case
\bea
\label{smpl}
&& \phi=0 \nn
&& G_{\mu\nu}=\delta_{\mu\nu} \nn
&& B_{mn}=B_{im}=B_{mi}=0\ .
\eea
Then we obtain,  
\be
\label{gfaction0}
S_D=k\int_0^T d\zeta_0 \int d^p\zeta 
\left[\det \left(\hat G_{ij}
+\partial_i X^\bot\cdot\partial_j X^\bot\right)
\right]^{1 \over 2}
\ee
and the effective potential is defined by 
\be
\label{EP}
V=-\lim_{T\rightarrow\infty}{1 \over T}
\ln \int {DX^\bot DA_i \over V_A}\e^{-S_D}\ .
\ee
Here $V_A$ is the gauge volume for the gauge field $A_i$ and 
\bea
\label{Xbot}
X^\bot &=& (X^{p+1},X^{p+2},\cdots,X^{d-1}) \\
\label{G}
\hat G_{ij}&=&R_iR_j\delta_{ij}+\cF_{ij} \nn
\label{cF} 
\cF_{ij}&\equiv&R_iR_jB_{ij}+F_{ij}\ .
\eea
As we can see from (\ref{G}), 
the anti-symmetric part $\cF_{ij}$ 
of $\hat G_{ij}$ contains the contribution from the 
anti-symmetric tensor $B_{ij}$ and the gauge field on 
the D-brane world sheet.
For a while, we treat the electromagnetic field 
strength $F_{ij}$ as classical field and assume 
$\cF_{ij}$ has a constant and nontrivial 
vacuum expectation value.

We now define the following tensors
\bea
\label{Gtensor}
G^{S\, ij}&\equiv&{1 \over 2} 
\left((\hat G^{-1})^{ij}+(\hat G^{-1})^{ji}
\right) \nn
\tilde G^{\alpha\beta}&\equiv&{1 \over G^{S\, 00}}\left(
G^{S\, \alpha\beta}-
{1 \over G^{S\, 00}}
G^{S\, 0\alpha}G^{S\, 0\beta}\right)
\eea
Here $(\hat G^{-1})^{ij}$ is the inverse matrix of $\hat G_{ij}$ and 
$\alpha,\beta=1,2,\cdots p$. 
Then the one-loop effective potential is given by (compare with \cite{12})
\be
\label{VT}
V_T=k(\det \hat G_{ij})^{1 \over 2}+ {1 \over 2}(d-p-1)
\sum_{n_1,n_2,\cdots , n_p=-\infty}^\infty
\left( 4\pi^2\sum_{k,l=1}^p \tilde G^{kl} n_k n_l 
\right)^{1 \over 2}\ .
\ee

In the following, we consider some examples for the choice of electromagnetic
background.

First we consider the membrane ($p=2$) with 
the magnetic background
\be
\label{mag}
\cF_{0k}=0\ ,\ \ \cF_{12}=-\cF_{21}=h 
\ee
and assume
\be
\label{R}
R_1=R_2=R\ .
\ee
Then we obtain
\bea
\label{dete}
\hat G&\equiv& \det \hat G_{ij}= R^4 + h^2 \\
\label{inv}
(\hat G^{-1})^{ij}&=&
\left(
\begin{array}{ccc}
1 & 0 & 0 \\
0 & \hat G^{-1}R^2 & -\hat G^{-1}h \\
0 & \hat G^{-1}h & \hat G^{-1}R^2 
\end{array} \right) \\
\label{Gtilde}
\tilde G^{\alpha\beta}&=&
\left(
\begin{array}{cc}
 \hat G^{-1}R^2 & 0 \\
 0 & \hat G^{-1}R^2 
\end{array} 
\right)
\eea
and we find that the one-loop potential has the following form
\be
\label{magV}
V_T=k(R^4+h^2)^{1 \over 2}+{d-3 \over 2}\cdot 
{ R \over ( R^4 + h^2 )^{1 \over 2}}f_T(1,1)\ .
\ee
Here we have used zeta-function regularization \cite{13} 
and we obtain \cite{12}
\be
\label{fT}
f_T(1,1)=2\pi\sum_{n_1,n_2=-\infty}^\infty \left(
n_1^2+n_2^2 \right)^{1 \over 2}=-1.438\cdots\ .
\ee
If we assume the magnetic flux 
\be
\label{flux}
\Phi= \int d\zeta_1 d\zeta_2 h
\ee
(note that $\cF_{ij}$ is a two-form) is size-independent, 
$h$ does not depend on $R$.
Contrary to $h=0$ case in \cite{12}, the effective potential 
(\ref{magV}) has a non-trivial minimum since 
$f_T(1,1)<0$. Hence, unlike the case $h=0$ toroidal D-brane
will not tend to collapse. There exists non-pointlike ground state
with finite radius. (The explicit value of this radius
is very complicated). This ground state is stabilized due to external
magnetic field effects.

In case that D-brane has open boundaries, we obtain the 
one-loop potential similar to (\ref{magV}).
The potential of the open D-brane has a non-trivial 
relative minimum besides $R=0$ when $k$ is small but $R=0$ becomes a 
stable minimum since $V_T(R>0)>V_T(R=0)$.

We also consider the one-loop potential of the membrane 
in the constant electric background where
\be
\label{ele}
\cF_{0k}=e\ ,\ \ \cF_{kl}=0\ ,\ \ R_1=R_2=R\ .
\ee
Then we find
\bea
\label{detG}
\hat G&\equiv& \det \hat G_{ij}=R^4 + 2e^2 R^2 \\
\label{inv2}
(\hat G^{-1})^{ij}&=& \hat G^{-1}
\left(
\begin{array}{ccc}
R^4 & -eR^2 & -eR^2 \\
eR^2 & R^2+e^2 & -e^2 \\
eR^2 & -e^2 & R^2+e^2 
\end{array}
\right) \\
\label{Ghat}
\tilde G^{kl}&=&\left(
\begin{array}{cc}
R^2+e^2 & - e^2 \\
-e^2 & R^2+e^2 
\end{array}
\right)\ .
\eea
The total one-loop potential is given by 
\bea
\label{eleV}
V_T&=&k(R^4+2e^2)^{1 \over 2} \nn
&& + {d-3 \over 2}\cdot 
{( R^2 + e^2 )^{1 \over 2} \over R^2}
\hat f_T(1,1, -{e^2 \over R^2 + e^2})\ .
\eea
Here $\hat f_T(1,1, s)$ is defined by using of modified Bessel 
function $K_\nu$ \cite{13}
\bea
\label{hatf}
\hat f_T(1,1, s)&\equiv& 2\pi \sum_{n_1,n_2=-\infty}^\infty
\left(n_1^2 + n_2^2 + 2s n_1 n_2 \right)^{1 \over 2} \nn
&=&4\pi\zeta(-1) - {\zeta(3) \over 2\pi} \Delta \nn
&& - 4 \Delta^{1 \over 2} \sum_{n=1}^\infty
{\sigma_2(n)\cos (2sn\pi) \over n}
K_1(\pi n \Delta^{1 \over 2})
\eea
and 
\be
\label{Delt}
\Delta\equiv 4-4s^2\ ,\ \ \sigma_s(n)\equiv \sum_{d|n} d^s\ .
\ee
In order to discuss the stability of the one-loop potential 
in Eq.(\ref{eleV}), 
we consider the behavior when $R\rightarrow 0$, 
i.e., $s\rightarrow -1$ and 
$\Delta\sim {4R^2 \over e^2}\rightarrow 0$ in Eq.(\ref{hatf}).
The asymptotic behavior of $\sigma_2(n)$ when 
$n\rightarrow \infty$ is given by,
\be
\label{sas}
\sigma_2(n)=n^2 + {\cal O}(n)\ . 
\ee
By using the expression for $K_1 (z)$
\be
\label{dBfn}
K_1 (z)=z \int_1^\infty dt \e^{-zt}(t^2-1)^{1 \over 2}
\ee
and Eq.(\ref{sas}), we find that the behavior of the third term 
in the r.h.s. of Eq.(\ref{hatf}) 
when $R\rightarrow 0$ is given by
\bea
\label{3hatf}
&&- 4 \Delta^{1 \over 2} \sum_{n=1}^\infty
{\sigma_2(n)\cos (2sn\pi) \over n}
K_1(\pi n \Delta^{1 \over 2}) \nn
&& \sim - 4 \pi \Delta
\int_1^\infty dt (t^2-1)^{1 \over 2}\sum_{n=1}^\infty n^2 
\left( \e^{-\pi \Delta^{1 \over 2} t}\right)^n \nn
&& \sim - 4\pi \Delta 
\int_1^\infty dt{ (t^2-1)^{1 \over 2} \over 
\left(1- \e^{-\pi \Delta^{1 \over 2} t}\right)^3} \nn
&& \sim - 4\pi^{-2} c_0 \Delta^{-{1 \over 2}} \nn
&& \sim -{2 c_0 e \over \pi^2 R} \ .
\eea
Here 
\be
\label{c_0}
c_0\equiv \int_1^\infty dt{ (t^2-1)^{1 \over 2} \over 
t^3}\ .
\ee
Eq.(\ref{3hatf}) tells that the potential (\ref{eleV}) 
appears to be unstable since 
it is unbounded below when $R\rightarrow 0$. It is not
strange because we expect that constant electrical field
(unlike the magnetic field) leads to pairs creation and
it should quickly destabilize the static approximation.
However, when non-perturbative effects are taken into account
(as we will see later for static potential in the same background)
they play more essential role near $R=0$ than destabilizing effect
of the electrical field. As a result static potential has stable
minimum even in the electrical background.

If $h$ in (\ref{mag}) and $e$ in (\ref{ele}) come 
from $B_{ij}$ in (\ref{cF}), 
they can be considered to be pure background fields and 
we need not to vary them to find the saddle point. 
Note that, however, they become size-dependent as found 
in (\ref{cF}) and we should replace them by
\be
\label{Rdep}
h\rightarrow R^2 h\ ,\ \ \ e\rightarrow R e\ .
\ee
The new parameters $h$ and $e$ are size-independent.

Even when $h$ and $e$ come from $F_{ij}$ in (\ref{cF}), 
we should not vary them since their flux would be 
conserved. If we assume the flux is conserved, $h$ and $e$ 
should be size-independent.

We can also consider the static potential in the limit 
of $d\rightarrow \infty$. 
By introducing the auxiliary fields $\sigma_{ij}$ and 
$\lambda^{ij}$, the action in (\ref{gfaction})  is rewritten 
 as
\bea
\label{gfla}
S_D&=&k\int_0^T d\zeta_0 \int d^p\zeta 
\Bigl[\left\{\det \left(\hat G_{ij} + \sigma_{ij}\right)
\right\}^{1 \over 2} \nn
&& -{k \over 2}\lambda^{ij}\left(
\partial_i X^\bot\cdot\partial_j X^\bot - \sigma_{ij}
\right)\Bigr]\ .
\eea
By integrating $X^\bot$ we obtain the 
large-$d$ effective action 
\bea
\label{Seff}
S_{eff}&=&{1 \over 2}(d-p-1)\Tr\ln
\left( -\partial_i \lambda^{ij}\partial_j\right) \ .
\nn
&&+kT\left[ \{\det(\hat G_{ij}+\sigma_{ij})\}^{1 \over 2}
-{1 \over 2} \Tr(\lambda\cdot\sigma) \right]
\eea
In the large $d$ limit, we can only consider the saddle 
point and the auxiliary fields $\sigma_{ij}$ and 
$\lambda^{ij}$ can be treated as classical fields.

We now consider the D-brane  with 
magnetic field background when $p$ is even 
\be
\label{mgbp}
f_{2n-1\,2n}=h\ , \hskip 1cm n=1,2,\cdots,{p \over 2}
\ee
in the large $d$ approximation. 
We assume 
\be
\label{RRRR}
R_i=R
\ee
and the auxiliary fields $\sigma_{ij}$ 
and $\lambda^{ij}$ have the following form in the 
saddle point
\bea
\label{diag}
\sigma_{ij} &=&{\rm diag}(\sigma_0, \sigma_1, \sigma_1) 
+ {\cal O}(1/d) \nn  
\lambda_{ij} &=&{\rm diag}(\lambda_0, \lambda_1, \lambda_1) 
+ {\cal O}(1/d) \ .
\eea
Since $\det(\hat G_{ij}+\sigma_{ij})$ in Eq.(\ref{Seff}) is 
given by
\be
\label{detGs}
\det(\hat G_{ij}+\sigma_{ij})=(1+\sigma_0)\left\{
(R^2+\sigma_1)^2 + h^2\right\}^{p \over 2}\ , 
\ee
we find that the effective action has the following form
\bea
\label{Seffd}
S_{eff}&=&kT\Bigl[(1+\sigma_0)^{1 \over 2}
\{(R^2+\sigma_1)^2 + h^2\}^{p \over 4} \nn
&& -{1 \over 2}R^2(\sigma_0\lambda_0+2\sigma_1\lambda_1) 
+ R^2\alpha \left({\lambda_1 \over \lambda_0}
 \right)^{1 \over 2}
\Bigr]
\eea
where $\alpha$ is defined by
\bea
\label{alpha}
\alpha&\equiv& {d-p-2 \over 2k R^{p+1}}f_T(1,1, \cdots, 1) \\
f_T(1,1, \cdots, 1)&\equiv& 2\pi
\sum_{n_1,n_2,\cdots,n_p=-\infty}^\infty 
\left( \sum_{i=1}^p  n_i^2
\right)^{1 \over 2} \ .
\eea
The variations with respect to $\sigma_0$,  
$\lambda_0$ and $\lambda_1$ give the following equations, 
respectively:
\bea
\label{s0eq}
R^p\lambda_0&=&(1+\sigma_0)^{-{1 \over 2}}
\{(R^2+\sigma_1)^2 + h^2\}^{p \over 4} \\
\label{l0eq}
\sigma_0&=&-\alpha \lambda_1^{1 \over 2}
\lambda_0^{-{3 \over 2}} \\
\label{l1eq}
p\sigma_1&=&\alpha \lambda_1^{-{1 \over 2}}
\lambda_0^{-{1 \over 2}} \ .
\eea
We now solve $\sigma_0$, $\lambda_0$, and $\lambda_1$ 
with respect to $\sigma_1$ by using Eqs.(\ref{s0eq}), 
(\ref{l0eq}) and (\ref{l1eq}).

By using (\ref{l0eq}) and (\ref{l1eq}), we obtain
\bea
\label{l0}
p\sigma_0\sigma_1&=&-\alpha^2\lambda_0^{-2} \\
\label{l1o0}
{\lambda_1 \over \lambda_0}&=&-{1 \over p}
{\sigma_0 \over \sigma_1} \\
\label{s0l0s1l1}
\sigma_0\lambda_0+p\sigma_1\lambda_1&=&0\ .
\eea
By using (\ref{s0eq}) and (\ref{l0}), we can solve for
$\sigma_0$:
\be
\label{s0}
\sigma_0=-{\alpha^2 R^{2p} \over 
\alpha^2 R^{2p} + p \sigma_1 \left\{ 
(R^2 + \sigma_1)^2 + h^2 \right\}^{p \over 2}}\ .
\ee
Then we find 
\bea
\label{Seff2}
S_{eff}&=&kT\left[{\sqrt p \over |\alpha| R^p}
\left\{(R^2 + \sigma_1)^2 +h^2 \right\}
(-\sigma_0)^{1 \over 2}\sigma_1^{1 \over 2}
+{1 \over \sqrt p}R^p|\alpha| {(-\sigma_0)^{1 \over 2} 
\over \sigma_1^{1 \over 2}}\right] \nn
&=&{kT \over \sqrt p}\sigma_1^{-{1 \over 2}}
\left[ 2\{(R^2+\sigma_1)^2 + h^2\}^{p \over 2}
\sigma_1 + \alpha^2 R^{2p}
\right]^{1 \over 2}
\eea
and the static potential $V_T$ with respect to 
$\sigma_1$ is given by
\be
\label{ldsp}
V_T={k \over \sqrt p}\sigma_1^{-{1 \over 2}}
\left[ 2\{(R^2+\sigma_1)^2 + h^2\}^{p \over 2}
\sigma_1 + \alpha^2 R^{2p}
\right]^{1 \over 2}\ .
\ee
Note that the static potential $V$ is defined by
\be
\label{sp}
V\equiv \lim_{T\rightarrow \infty} {1 \over T}\cdot 
S_{eff}\ .
\ee
We can easily find that there exits a stable 
non-trivial minimum in Eq.(\ref{ldsp}).
The minimum exists even when $h=0$ in \cite{12}.

The large $d$ effective potential 
in the electric background:
\be
\label{ele2}
\cF_{0k}=e_k\ ,\ \ \cF_{kl}=0\ ,\ \ R_k=R
\ee
can be found similarly.
Since we find 
\be
\label{detGs2}
\det(\hat G_{ij}+\sigma_{ij})=(1+\sigma_0)(R^2+\sigma_1)^p 
+e^2(R^2+\sigma_1)^{p-1}
\ee
where
\be
\label{e2}
e^2\equiv \sum_{k=1}^p e^2_k\ .
\ee
The effective action is given by
\bea
\label{Seffd2}
S_{eff}&=&kT\Bigl[\{(1+\sigma_0)(R^2+\sigma_1)^p 
+e^2(R^2+\sigma_1)^{p-1}
\}^{1 \over 2} \nn
&& -{1 \over 2}R^p(\sigma_0\lambda_0+p\sigma_1\lambda_1) 
+ R^p\alpha \left({\lambda_1 \over \lambda_0}
 \right)^{1 \over 2}
\Bigr]
\eea
Here $\alpha$ is defined in (\ref{alpha}).
The variations with respect to $\sigma_0$, 
$\lambda_0$ and $\lambda_1$ give the following equations, 
respectively:
\bea
\label{s0eq2}
R^p\lambda_0&=&\{(1+\sigma_0)(R^2+\sigma_1)^p 
+e^2(R^2+\sigma_1)^{p-1}
\}^{-{1 \over 2}}(R^2+\sigma_1)^p \\
\label{l0eq2}
\sigma_0&=&-\alpha \lambda_1^{1 \over 2}
\lambda_0^{-{3 \over 2}} \\
\label{l1eq2}
p\sigma_1&=&\alpha \lambda_1^{-{1 \over 2}}
\lambda_0^{-{1 \over 2}} \ .
\eea
By using (\ref{l0eq2}) and (\ref{l1eq2}), we obtain
\bea
\label{l02}
p\sigma_0\sigma_1&=&-\alpha^2\lambda_0^{-2} \\
\label{l1o02}
{\lambda_1 \over \lambda_0}&=&-{1 \over p}
{\sigma_0 \over \sigma_1} \\
\label{s0l0s1l12}
\sigma_0\lambda_0+p\sigma_1\lambda_1&=&0\ .
\eea
By using (\ref{s0eq2}) and (\ref{l02}), we can solve for 
$\sigma_0$:
\be
\label{s02}
\sigma_0=-{\alpha^2 R^{2p}
\left\{1+e^2(R^2 + \sigma_1)^{-1}\right\} 
\over \alpha^2 R^{2p} + p \sigma_1 (R^2 + \sigma_1)^p}\ . 
\ee
Then we find 
\bea
\label{Seff22}
S_{eff}&=&kT\left\{{\sqrt p \over \alpha R^p}
(R^2 + \sigma_1)^p 
(-\sigma_0)^{1 \over 2}\sigma_1^{1 \over 2}
+{1 \over \sqrt p}R^p\alpha {(-\sigma_0)^{1 \over 2} 
\over \sigma_1^{1 \over 2}}\right\} \nn
&=&{kT \over \sqrt p}\sigma_1^{-{1 \over 2}}
\left\{1+e^2(R^2 + \sigma_1)^{-1}\right\}^{1 \over 2}
\left[ p(R^2+\sigma_1)^p \sigma_1 + \alpha^2 R^{2p}
\right]^{1 \over 2} \\
\label{eplde}
V_T&=&{k \over \sqrt p}\sigma_1^{-{1 \over 2}}
\left\{1+e^2(R^2 + \sigma_1)^{-1}\right\}^{1 \over 2}
\left[ p(R^2+\sigma_1)^p \sigma_1 + \alpha^2 R^{2p}
\right]^{1 \over 2} .
\eea
The static potential (\ref{eplde}) has also only 
one stable minimum.
The minimum also exists when $e=0$.

Hence, one can see that non-perturbative effects of large d
expansion play more essential role than destabilizing effect
of the constant electrical field. As a result this potential
shows the possibility of non-trivial minimum (non-pointlike
ground state).

\section{One-loop potential with the quantum back-reaction of  
electromagnetic field.}

In the following, we consider the quantum back-reaction 
of the gauge field 
by dividing the anti-symmetric part $\cF_{ij}$ in 
$\hat G_{ij}$ into the sum of 
the classical part $\cF_{ij}^c$ and 
quantum fluctuation $F_{ij}^q$:
\be
\label{div}
\cF_{ij}=\cF_{ij}^c + F_{ij}^q\ . 
\ee
In the following, we write $R_i^2\delta_{ij}+\cF_{ij}^c$ 
as $\hat G_{ij}$ and $F_{ij}^q$ as $F_{ij}$.

We use the one-loop approximation by expanding the action 
(\ref{gfaction}) and keeping the quadratic term.
Since 
\bea
\label{der1}
{\partial \hat G \over \partial \hat G_{ij}}&=&\hat G(\hat G^{-1})^{ji} \\
\label{der2}
{\partial^2 \hat G \over \partial \hat G_{ij}
\partial \hat G_{kl}}&=&\hat G\left\{-(\hat G^{-1})^{jk}(\hat G^{-1})^{li}
+(\hat G^{-1})^{ji}(\hat G^{-1})^{lk} \right\}
\eea
we find
\bea
\label{dsqrt}
&&\left\{\det \left(\hat G_{ij}+F_{ij}\right)
\right\}^{1 \over 2} \nn
&& =\hat G^{1 \over 2}
\Bigl[ 1 + {1 \over 2}(\hat G^{-1})^{ji}F_{ij} \nn
&& +\left\{ -{1 \over 4}(\hat G^{-1})^{jk}(\hat G^{-1})^{li}
+{1 \over 8}(\hat G^{-1})^{ji}(\hat G^{-1})^{lk} 
\right\}F_{ij}F_{kl} + \cdots \Bigr] \ .
\eea
If $\hat G_{ij}$ is a constant tensor, 
the second term is a total derivative and does not 
give any contribution. 

Let's consider the following easy example, 
where $\hat G_{ij}$ is given by
\be
\label{fG}
\hat G_{ij}=R_i^2\delta_{ij}
\ee
Then
\be
\label{fdet}
\left\{\det \left(\hat G_{ij}+F_{ij}\right)
\right\}^{1 \over 2} 
=\hat G^{1 \over 2}
\left[ 1 -
{1 \over 4}\sum_{i,j=0}^p {1 \over R_i^2 R_j^2}
F_{ij}^2 + \cdots \right] \ .
\ee
Here we neglect total derivative terms.
By rescaling the gauge potential 
\be
\label{rescl}
A_i\rightarrow R_i A_i
\ee
($R_0=1$), 
Eq.(\ref{fdet}) is rewritten by
\bea
\label{fdet2}
\left\{\det \left(\hat G_{ij}+F_{ij}\right)
\right\}^{1 \over 2} 
&=&\hat G^{1 \over 2}
\Bigl[ 1 +
{1 \over 2}\sum_{i,j=0}^p A_i\left\{\left(
\sum_{k=0}^p{\partial_k^2 \over R_k^2 }\right)
\delta^{ij}- {\partial_i \partial_j \over R_i R_j}
\right\}A_j  \nn
&& + \mbox{total derivative terms} + 
\cdots \Bigr] \ .
\eea
If we choose the gauge condition 
\be
\label{gcon}
A_0=0
\ee
we find the residual gauge condition (transverse 
condition)
\be
\label{trcon}
\sum_{i=1}^p{1 \over R_i}\partial_i A_i =0 
\ee
and we obtain
\bea
\label{fdet3}
&&\left\{\det \left(\hat G_{ij}+F_{ij}\right)
\right\}^{1 \over 2} \nn
&&=\hat G^{1 \over 2}
\left[ 1 +
{1 \over 2}\sum_{\begin{array}{ll} i: &\mbox{\footnotesize 
transverse} \\ & \mbox{\footnotesize components}
\end{array}} A_i\left(
\sum_{k=0}^p{\partial_k^2 \over R_k^2 }\right)
A_i  + \cdots \right] \ .
\eea
Then we find that the contribution 
from the gauge field to the one-loop potential is given by
\be
\label{excon}
{1 \over 2}(p-1)\sum_{n_1,n_2,\cdots,n_p=-\infty}^\infty 
\left( \sum_{i=1}^p{4\pi^2 n_i^2 \over R_i^2}
\right)^{1 \over 2}\ .
\ee
Therefore the qualitative nature of the one-loop potential 
does not change if compare 
to the case without the contribution from the gauge field.

The one-loop potential in the toroidal 
membrane ($p=2$) with the magnetic background 
(\ref{mag})  (including the quantum backreaction of the 
gauge fields)  can be obtained similarly.
The kinetic term of the gauge potential is 
given by
\bea
\label{kterm}
&& \left\{ -{1 \over 4}(\hat G^{-1})^{jk}(\hat G^{-1})^{li}
+{1 \over 8}(\hat G^{-1})^{ji}(\hat G^{-1})^{lk} 
\right\}F_{ij}F_{kl} \nn
&& = {1 \over 2}\hat G^{-1}R^2 \left(
F_{01}^2 + F_{02}^2 \right) 
+{1 \over 2}\hat G^{-2}R^4 F_{12}^2\ .
\eea
By choosing the gauge fixing condition (\ref{gcon}), 
we obtain the transverse condition 
\be
\label{trcon2}
\partial_1 A_1 + \partial_2 A_2 =0\ .
\ee
By using (\ref{gcon}) and (\ref{trcon2}) and rescaling 
the gauge potential 
\be
\label{resale}
A_{1,2}\rightarrow \hat G^{1 \over 2}R^{-1}A_{1,2}\ ,
\ee 
Eq.(\ref{kterm}) 
can be rewritten as
\bea
\label{kterm2}
&& \left\{ -{1 \over 4}(\hat G^{-1})^{jk}(\hat G^{-1})^{li}
+{1 \over 8}(\hat G^{-1})^{ji}(\hat G^{-1})^{lk} 
\right\}F_{ij}F_{kl} \nn
&& ={1 \over 2}\sum_{i=1,2}
\left( \partial_0 A_i \right)^2
+{1 \over 2}\cdot {R^2 \over \hat G}
\sum_{\alpha,\beta=1,2}\left( \partial_\alpha A_\beta 
\right)^2\ .
\eea
This tells that the contribution to the one-loop potential from 
the gauge potential is given by
\be
\label{epot}
{1 \over 2}\cdot {R \over (R^4 + h^2)^{1 \over 2} } f_T(1,1)\ .
\ee
Then the total one-loop potential is given by 
\be
\label{magV2}
V_T=k(R^4+h^2)^{1 \over 2}+{d-2 \over 2}\cdot 
{ R \over ( R^4 + h^2 )^{1 \over 2}}
f_T(1,1)\ .
\ee
When we compare this potential with that of Eq.(\ref{magV}), 
the only difference is the coefficient in the second term.
Therefore the potential (\ref{magV2}) is stable and has non-trivial minimum as in Eq.(\ref{magV}).

This potential appears to be unstable since 
it is unbounded below when $R\rightarrow 0$.
However, it would not be true since the non-perturbative
effect would play more essential role near $R=0$.
Hence, the quantum effects of abelian gauge field
tend to destroy the ground state which exists when
D-brane interacts with constant magnetic field.

The one-loop potential in the toroidal 
membrane ($p=2$) with the electric background when  
including the quantum backreaction of the 
gauge potential can be also found.
The kinetic term of the gauge potential is 
given by
\bea
\label{kterm22}
&& \left\{ -{1 \over 4}(\hat G^{-1})^{jk}(\hat G^{-1})^{li}
+{1 \over 8}(\hat G^{-1})^{ji}(\hat G^{-1})^{lk} 
\right\}F_{ij}F_{kl} \nn
&& = {1 \over 2}R^4 \hat G^{-1}(\hat G^{-1})^{\alpha\beta}
F_{0\alpha}F_{0\beta} 
+ {1 \over 4}(\hat G^{-1})^{\alpha\beta}(\hat G^{-1})^{\gamma\delta}
F_{\alpha\gamma}F_{\beta\delta} \ .
\eea
By choosing the gauge condition (\ref{gcon}),
we obtain the transverse condition 
\be
\label{trcon3}
(\hat G^{-1})^{\alpha\beta}\partial_\alpha A_\beta=0\ .
\ee
By using (\ref{gcon}) and (\ref{trcon3}), 
Eq.(\ref{kterm22}) 
can be rewritten as
\bea
\label{kterm23}
&& \left\{ -{1 \over 4}(\hat G^{-1})^{jk}(\hat G^{-1})^{li}
+{1 \over 8}(\hat G^{-1})^{ji}(\hat G^{-1})^{lk} 
\right\}F_{ij}F_{kl} \nn
&& ={1 \over 2} \hat G^{-1}R^4 (\hat G^{-1})^{\alpha\beta}
\partial_0 A_\alpha \partial_0 A_\beta \nn
&& +{1 \over 2} (\hat G^{-1})^{\alpha\beta}(\hat G^{-1})^{\gamma\delta}
\partial_\gamma A_\alpha \partial_\delta A_\beta \ .
\eea
Then the contribution to the one-loop potential from 
the gauge potential is given by
\be
\label{epot2}
{1 \over 2}\cdot {(R^2 + e^2)^{1 \over 2} 
\over R^2 } 
\hat f_T(1,1, -{e^2 \over R^2 + e^2})\ .
\ee
Then the total one-loop potential is given by 
\bea
\label{eleV2}
V_T&=&k(R^4+2e^2)^{1 \over 2}
+{d-2 \over 2}\cdot {(R^2 + e^2)^{1 \over 2} \over R^2}
\hat f_T(1,1, -{e^2 \over R^2 + e^2})\ .
\eea
This potential also appears to be unstable since 
it is unbounded below when $R\rightarrow 0$.

Similarly, one can find the contribution of gauge 
fields to static
potential. However, in this case such contribution 
is next-to-leading
order of $1/d$-expansion. Hence, it is not relevant 
for study of extremum of static potential.

\section{Discussion}

In the present work we discussed the one-loop 
and static potentials
for toroidal D-brane in constant electromagnetic field. 
The interesting qualitative result of such study 
is the fact that classical electromagnetic
background may stabilize the one-loop potential. 
As a result the non-trivial
minimum of potential exists (i.e. non-pointlike ground 
state of D-brane exists).

The properties of static potential (where quantum effects 
of gauge field
are next-to-leading order) are described in detail. 
It is found that 
in all cases static potential has stable minimum.

It is not difficult to generalize the results of 
this study to other backgrounds. For example, one can 
consider the spherical D-brane in constant electromagnetic field. 
Here, we found very similar results for one-loop 
and static potentials
as in above case of toroidal D-brane.

Note that we worked in the static approximation.
However, it is well-known fact that static approximation 
is quickly broken in the external electrical field where 
intensive pairs production occurs.
Hence, as extension of this work it would be very interesting 
to calculate
the effective action for D-brane in an external electrical field
taking into account also imaginary part of the effective action,
i.e. to work beyond static approximation.

Another interesting problem is to study the effective potential
for D-brane described by non-abelian DBI action in the
background gauge fields. We hope to return to this problem in future research.

\ 

\noindent
{\bf\large Acknowledgement}

We are very indebted to A. Sugamoto for discussions.

\end{document}